**Unveiling the photophysics of thiourea from CASPT2/CASSCF potential energy surfaces and singlet/triplet excited state molecular dynamics simulations.**

Neus Aguilera Porta,[a] Giovanni Granucci*[,b], Jordi Munoz-Muriedas[c] and Inés Corral*[a,d]

[a] Departamento de Química, Universidad Autónoma de Madrid, C/ Francisco Tomás y Valiente 7, 28049 Cantoblanco, Madrid, Spain.

[b] Università di Pisa, Dipartimento di Chimica e Chimica Industriale, Via Giuseppe Moruzzi, 13, 56124 Pisa, Italy.

[c] Computational Toxicology, GlaxoSmithKline, Park Rd Ware, Hertfordshire, UK

[d] IADCHEM. Institute for Advanced Research in Chemistry, Universidad Autónoma de Madrid, 28049 Cantoblanco, Madrid, Spain

# Abstract

This work describes the decay mechanism of photoexcited thiourea, both in gas phase and in solution, from the information inferred from the topography of the excited and ground state potential energy surfaces and mixed singlet/triplet quantum classical molecular dynamics simulations. Our gas phase results reveal $T_1/S_0$ intersystem crossing as the dominant (49%) intrinsic decay channel to the ground state, which reaches a population of 0.28 at the final time of our simulations (10 ps). Population of the $T_1$, would occur after internal conversion to the $S_1$ from the spectroscopic $S_2$ electronic state, followed by $S_1$->$T_2$ intersystem crossing and $T_2$->$T_1$ internal conversion processes. Minor decay channels occurring exclusively along the singlet manifold, i.e. $S_2$->$S_0$ (33%) and $S_1$->$S_0$ (18%), were also observed to play a role in the relaxation of photoexcited thiourea in the gas phase. The explicit incorporation of water-thiourea interactions in our simulations was found to provoke a very significant delay in the decay to the ground state of the system, with no transitions to the $S_0$ being registered during the first 10 ps of our simulations. Intermolecular vibrational energy redistribution and explicit hydrogen bond interaction established between water molecules and the $NH_2$ group of thiourea were found to structurally or energetically hamper the access to the intersystem crossing or internal conversion funnels with the $S_0$.

# Introduction

Thioureas are the sulfur analogues of the organic compounds ureas, which result from the exchange of the carbonyl group by the thiocarbonyl function. Despite their structural resemblance, the very different electronegativity of oxygen and sulfur, along with the lower ionization potential and excitation energies of S containing species compared to their oxygen counterparts confer these two families of compounds significantly different physico-chemical properties. For instance, thiourea is a much stronger acid than urea (p$K_A$ = 21.1 and 26.9, respectively in DMSO) [1]. This has been, for instance, wisely profited in the design of recognition receptors for specific anions [2]. But the differences between urea and thiourea are not only limited to ground state properties. Indeed, the exchange of C=O by C=S is also expected to induce changes in the optical properties and the photophysics of these systems. Interestingly, the third harmonic generation value estimated for thiourea was found to be one order of magnitude larger than in urea, suggesting thiourea crystals as very attractive materials for the study of non-linear optical properties[3]. Carbonyl-by-thiocarbonyl substitution has also been found to dramatically change the topography of the excited state potential energy surfaces of other systems. This is the case of thionucleobases, [4] [5] which on the one hand show strongly red shifted absorption spectra compared to their canonical counterparts and for which triplet quantum yields close to unity have been registered. These changes in the absorption spectrum and in the photophysical properties of thiosubstituted nucleobase analogues have been ascribed, on the one hand, to the stabilization of the excitations localized on the thiocarbonyl groups and, on the other, to the particular shape of the excited potential energy surfaces of these systems that would favor the trapping of the population in singlet excited minima, in the presence of close lying triplet states. This, added to the heavy atom effect introduced by the sulfur atom in these molecules, would favor the transfer of population to the triplet manifold.

The relaxation mechanism from the first excited singlet of urea, was outlined by Fang and coworkers some years ago [6], in a study focused on the investigation of the mechanism behind its photochemical deamination and dehydration, experimentally observed upon irradiation of the system with light of 160 nm wavelength. [7] These authors propose two possible routes for the deactivation of photoexcited urea to the ground state. The first involves direct relaxation through a $S_1/S_0$ funnel, whilst the second involves the deactivation of the system through the population of triplet intermediate states. These two decay funnels were found to be energetically accessible from the $S_1$ minimum, considering the Franck-Condon $S_1$ initial energy of the system ($E_{S1FC}$: 6.87 eV, $E_{S1min}$: 4.64 eV, $E_{S1/T1}$: 4.77 eV, $E_{S1/S0}$: 4.66 eV). The vibrationally hot population reaching the ground state via the $S_1/S_0$ funnel would easily undergo deamination or dehydration processes following different H transfer reactions. In this way, $NH_3$ loss would occur after H migration from one $NH_2$ group to the other. Alternatively, $NH_3$ formation would take place from the reallocation of the H from the NH moiety to the $NH_2$ group at the $NH_2C(OH)NH$ intermediate arising itself from an amino to carbonyl H transfer. The same intermediate would precede the loss of water if the transfer of the second H atom involves instead the NH and OH groups. The authors map similar pathways for deamination and dehydration reactions along the $T_1$ potential, though involving higher energy barriers, thus, not being competitive with the ground state processes.

Similarly to ureas, thiourea derivatives have a great importance in agriculture, where they are extensively employed as plants and insects growth regulators, or for the purpose of killing fungus spores and weed [8]. Several studies have demonstrated, however, that plant growth controllers are sensitive to the effect of light [9].

This paper provides a comprehensive overview of the intrinsic potential deactivation routes of photoexcited thiourea, based on quantum chemical calculations, and their actual competition for the molecule in the gas phase and immersed in a water cluster after the analysis of the results from molecular dynamics simulations. Both the static and time resolved picture of the relaxation mechanism of photoexcited thiourea provided by our results would help rationalizing the effect produced by carbonyl-by-thiocarbonyl substitution in urea.

# Computational Details

Thiourea ground state equilibrium geometry was optimized at the MP2/6-311G(d,p) level of theory. [10], [11] [12]

The CASPT2 [13] [14] absorption spectrum up to 8 eV was simulated considering a 5 roots State Average (SA) CASSCF [15] reference wave function constructed with an active space comprising 10 electrons distributed into 7 orbitals, (see Figure 1). This includes the complete set of π orbitals, one lone pair from the C=S group and the pair of bonding and antibonding σ orbitals from the thiocarbonyl group. For these calculations, the ANO-L basis set, [16, 17] with a H [3s2p], C [4s3p2d], N [4s3p2d], S [5s4p2d] contraction was employed.

Starting from the lowest lying spectroscopic state at the Franck Condon (FC) region, the topography of the singlet and triplet excited state potential energy surfaces was explored with the help of minimum energy path (MEP) calculations[18]. For this purpose, we resorted to the SA(3)-CASSCF protocol, with the active space described above. Here, the ANO-S basis set, [19] with a H [2s1p], C [3s2p1d], N [3s2p1d], S [4s3p2d] contraction was used.

Singlet and triplet multiplicity minima were further reoptimized at the same level of theory, with tighter convergence criteria.

Singlet/singlet, singlet/triplet and triplet/triplet conical intersections were initially located at the CASSCF/6-31G* [15] [20, 21] level of theory, and then reoptimized with the small atomic natural orbital basis.

Final energies for all the minima, and the crossing points, were calculated with MS(3)-CASPT2//SA(3)-CASSCF and the ANO-L basis set. A level shift [22] of 0.3 a.u. and the default IPEA [23] shift of 0.25 a.u. were employed throughout the calculations.

For those crossings where the incorporation of dynamical correlation was found to lift the interstate degeneracies calculated at the CASSCF level of theory, new geometries were located following CASPT2 MEPs, using SA(3)-CASSCF(10,7)/ANO-S wavefunctions as a reference.

Spin-orbit couplings (SOC) were calculated with the relativistic one electron effective Douglas-Kroll-Hess Hamiltonian [24-27] and the Atomic Mean Field Integral approximation[28, 29] . For the relativistic calculations, the ANO-RCC basis set contracted as, H [2s1p], C [3s2p1d], N [3s2p1d], S [4s3p2d] was used instead.[30] [31]

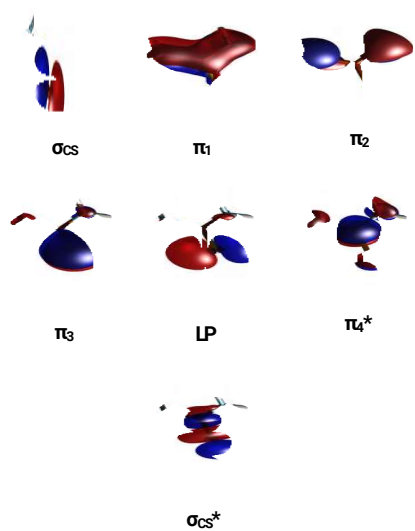

**Figure 1**.- Active space considered for the calculation of the vertical absorption spectrum.

Minima optimization at the CASSCF level of theory and final CASPT2 single point calculations were performed with MOLCAS 7.6 [32, 33] Singlet/singlet and singlet/triplet conical intersections were located with MOLPRO 2009[34].

The decay dynamics of photoexcited thiourea was studied with the mixed quantum-classical surface hopping algorithm, [35] adopting the adiabatic representation.[36] During the integration of the nuclear trajectories, the energies, gradients and couplings were evaluated on the fly, in the framework of the semi-empirical AM1 configuration interaction with floating occupation molecular orbitals (FOMO-CI) approach [37], [38]. The wavefunctions were of CAS-CI type, with an active space of 4 electrons in 3 MOs. In particular, the orbitals included in the active space were the C=S lone pair, $\pi_3$ and a $\pi_4^*$.

The spin-orbit coupling between FOMO-CI wavefunctions was computed using a mean field Hamiltonian [39], with the relevant semiempirical spin-orbit parameters set to 28.6, 4 and 500 cm$^{-1}$ for C, N and S, respectively . The first three singlets and two triplets were considered in the dynamics simulations, and the same approach described in reference [40] was used (including the treatment of quantum decoherence). Two sets of simulations were performed: with and without solvent. In the former case, thiourea was inserted in a spherical cluster of 777 water molecules, using a QM/MM scheme with electrostatic embedding and the TIP3P force field for water molecules. Evaporation of water molecules from the surface of the sphere was avoided by adding a confining potential [41]. The initial conditions for the surface hopping trajectories were sampled from thermally equilibrated trajectories, run for 100 ps long at the temperature of 300 K for thiourea in vacuo and in the solvent cluster, using the Bussi-Parrinello algorithm [42]. The sampling procedure takes into account radiative transition probabilities as described in refs. [38, 39]. The nonadiabatic trajectories were run for a total of 10 ps, with a time step $\Delta t = 0.1$ fs. For the vacuum simulations, a total of 503 trajectories were run and 58 were interrupted for technical reasons, mainly due to orbitals flipping in and out from the active space, leading to discontinuities in the PESs and therefore in the total energy (the threshold for total energy conservation was set to 0.002 au). The analysis of the results was then performed on 445 trajectories. The solvent cluster simulations were done on a batch of 302 trajectories, of which only one was discarded for technical reasons.

# Results and discussion

Similarly to urea, [6],[43] two minima were located along the ground state MP2/6-311G(d,p) potential energy surface, See Figure 2. These two minima, with $C_s$ and $C_2$ symmetries, only differ in the internal dihedral angles of the $NH_2$ groups and were found to be almost energetically degenerate, the $C_2$ being slightly more stable (ca. 1 Kcal mol$^{-1}$). A $C_1$ symmetry transition state connecting the two minima was also located 0.69 eV over the $C_2$ minimum.

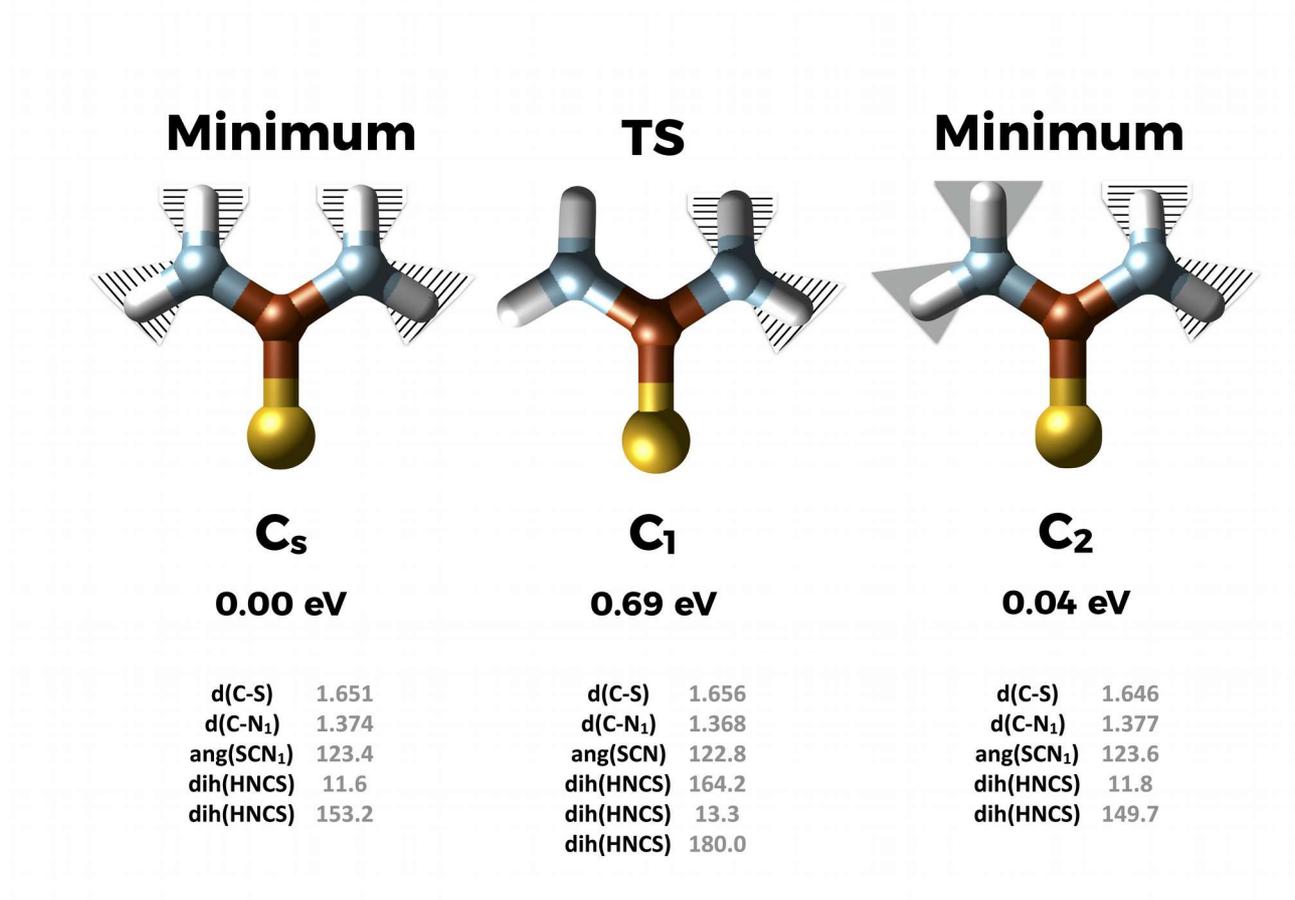

**Minimum**      **TS**      **Minimum**

$C_s$      $C_1$      $C_2$

0.00 eV      0.69 eV      0.04 eV

| | | | | | | | |
|---|---|---|---|---|---|---|---|
| d(C-S) | 1.651 | | d(C-S) | 1.656 | | d(C-S) | 1.646 |
| d(C-N$_1$) | 1.374 | | d(C-N$_1$) | 1.368 | | d(C-N$_1$) | 1.377 |
| ang(SCN$_1$) | 123.4 | | ang(SCN) | 122.8 | | ang(SCN$_1$) | 123.6 |
| dih(HNCS) | 11.6 | | dih(HNCS) | 164.2 | | dih(HNCS) | 11.8 |
| dih(HNCS) | 153.2 | | dih(HNCS) | 13.3 | | dih(HNCS) | 149.7 |
| | | | dih(HNCS) | 180.0 | | | |

**Figure 2.-** Optimized geometries of the ground state $C_s$ and $C_2$ minima of thiourea and the TS interconnecting them. Distances in Å and bond angles and dihedrals in degrees. Relative energies in eV.

The gas phase calculated absorption spectra for the most stable C$_2$ conformer, shown in Figure 3, shows an intense absorption, peaking around 5.50 eV, ascribed to $\pi_3$->$\pi_4$* excitation.

**Table 1.-** Thiourea vertical excitation energies computed at MS-CASPT2/CASSCF(10,7)/ANO-L and AM1 (normal font and italics, respectively). Urea vertical excitation energies, extracted from ref [6] and calculated at EOM-CCSD/6-311++G level of theory, are provided within parenthesis.

| | State Character | $\Delta E(eV)$ | | | $\Delta E(nm)$ | $f$ |
|---|---|---|---|---|---|---|
| **S$_1$** | LP→π$_4$* | 4.21 | *2.45* | (6.38) | 294 *506* (194) | 0.0003 (-) |
| **S$_2$** | π$_3$→π$_4$* | 5.50 | *3.90* | (6.51) | 225 *318* (190) | 0.5581 (-) |
| **S$_3$** | π$_2$→π$_4$* | 6.84(-) | | (-) | 181 (-) | 0.0304 (-) |
| **S$_4$** | LP→σ* | 7.76 (-) | | (-) | 160 (-) | 0.0173 (-) |

This transition is preceded by a weak n$\pi_4$* transition peaking at 4.21 eV. At higher energies (ca. 7 eV), we register another transition involving the redistribution of electrons along the $\pi$ cloud ($\pi_2$->$\pi_4$*) followed by an n$\sigma$* transition peaking at ca. 7.8 eV. These calculated transitions are in reasonable agreement with the experimental spectra recorded in diethylether, acetonitrile and ethanol, which locate the first two bands in the region of 4.3-4.4 eV and 4.9-5.1 eV.

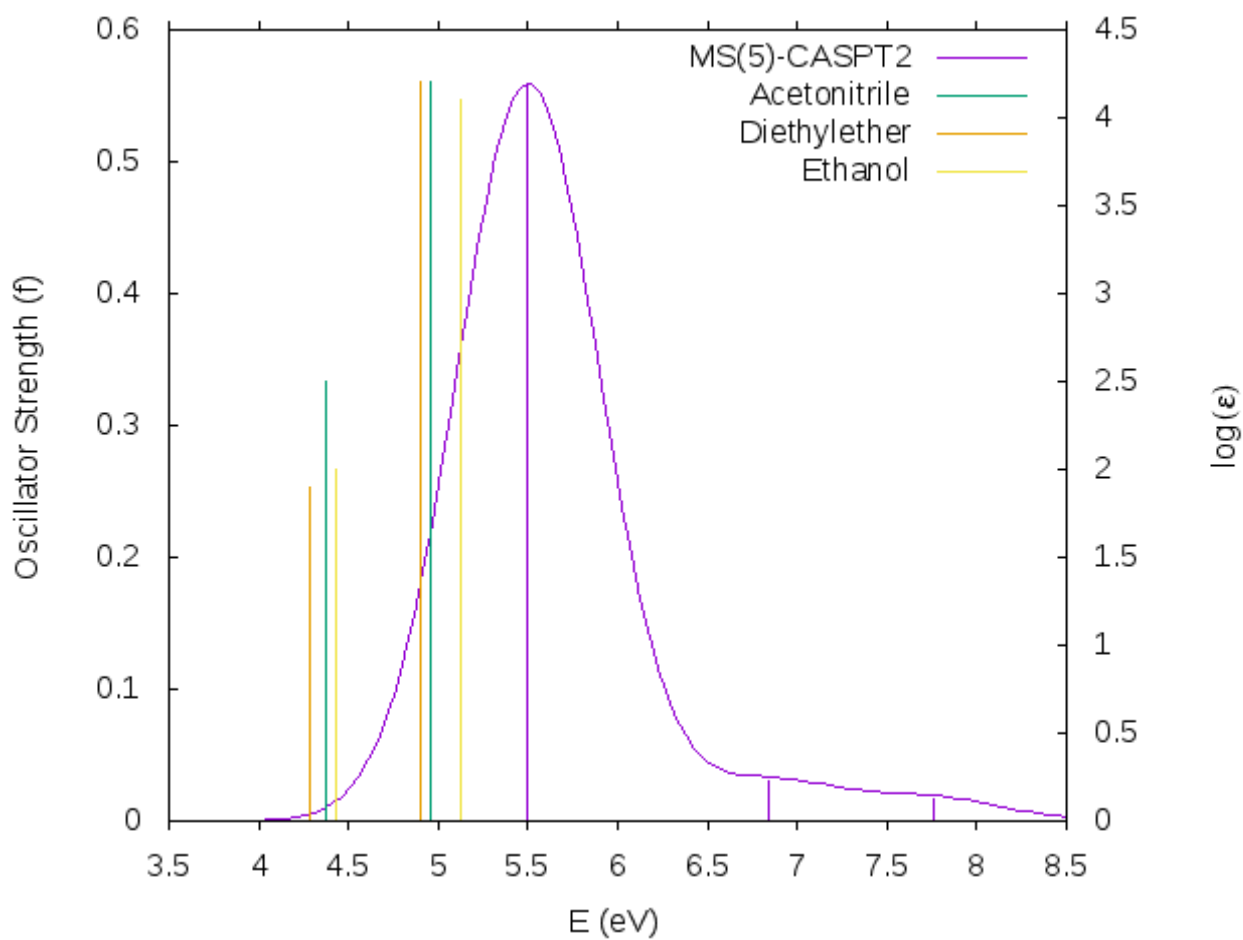

**Figure 3**.- Thiourea computed and experimental absorption spectra (Energies in eV). Computed excitation energies at MS-CASPT2/CASSCF(10,7)/ANO-L level of theory. Experimental absorption extracted from Ref [44]. Red, green and blue lines denote experiments performed in Acetonitrile, Diethylether, and Ethanol.

As already reported in other works [45], [46], [47], [48] carbonyl by thiocarbonyl substitution shifts to lower energies the electronic excitations involving this group, as a result of the stabilization of the thiocarbonyl molecular orbitals. Thus, although the lowest energy part of the absorption spectrum of both urea and thiourea consist of the same electronic transitions,[6],[49],[50] the absorption spectrum of the latter is red shifted by almost 1 eV. This shift, however, is not homogeneous along the whole spectrum. Whilst in urea the energy gap between the first two bands amounts to 0.1 eV, the separation between the $S_1$ and $S_2$ transitions in thiourea grows by more than 1 eV, compare thiourea and urea vertical excitation energies in Table 1.

Figure 4 schematically summarizes the excited state potential energy profile of thiourea relevant to its deactivation mechanism from the bright state. Upon irradiation of the system with light of 238 nm wavelength, the FC spectroscopic state, $S_2$, with $\pi_3 \rightarrow \pi_4^*$ character is expected to populate. Following the minimum energy path, the excited population would then relax to a minimum in this potential, $S_{2min}$, located 4 eV above the GS most stable minimum. Relaxation to the $S_{2min}$ provokes the stretching of the CS bond simultaneous to the pyramidalization of the C atom, reducing the symmetry of the system to $C_1$, see Table S1. The closest triplets to the $S_{2min}$, $T_1$ and $T_2$, respectively lie at 3.51 eV and 3.56 eV, that is 0.5 eV below the minimum, whilst the energy gap with the triplet $T_3$, located at 6.32 eV, is significantly larger. Further decay to lower lying electronic states is possible via the internal conversion funnel ($S_2/S_1$), located 0.23 eV over $S_{2min}$, but still well below the initial energy of the system. Accessing this funnel involves further stretching of the C-S bond relative to the $S_{2min}$ and the simultaneous planarization and rotation of the two $NH_2$ groups, leading to an almost planar $C(NH_2)_2$ moiety, recall Table S2. Once in the $S_1$ potential, the system is expected to reach the $S_{1min}$ (3.24 eV), with a reinforced CS bond compared to the $S_{2min}$ and presenting rotated $NH_2$ groups. A new conical intersection, involving the $S_1$ and the $S_0$, and, thus, allowing the decay of the population to the ground state, was located ca. 0.6 eV over the minimum. Relaxation to the ground state requires an important structural rearrangement of the system, imposing the $C(NH_2)_2$ moiety to lie almost perpendicular to the CS axis, see Table S1 and S2. Interestingly, the energy gaps with the triplets at the position of the $S_1$ minimum are much smaller than for the $S_{2min}$ (the $T_1$ and $T_2$ lie, respectively, only 0.1 eV below and above the minimum). In fact, we were able to locate a minimum energy $S_1/T_2$ crossing in the vicinity of the $S_{1min}$, lying 3.29 eV above the ground state. The complementary characters of the singlet $S_1$ ($n \rightarrow \pi_4^*$) and the triplet $T_2(\pi_3 \rightarrow \pi_4^*)$ transitions, both localized on the thiocarbonyl group, forecast a strong coupling between the $S_1$ and the $T_2$ at the position of this minimum, according to the qualitative El Sayed rules [51]. In fact, a spin orbit coupling (SOC) value of 285 cm$^{-1}$ was calculated at this region of the PES. The minimum energy path from $S_1/T_2$ crossing along the triplet manifold leads to a minimum in the $T_2$ potential. Further decay along the triplet manifold from the $T_{2min}$ is possible after accessing the $T_2/T_1$ degeneracy region located ca. 0.1 eV above the minimum, allowing the population of the most stable triplet minimum of $n \rightarrow \pi^*_4$ character. Similarly to urea [6], the structure of this triplet minimum reminds very much the one of the $S_{1min}$. Non-radiative relaxation from the $T_1$ is possible via the $T_1/S_0$ ISC funnel, located only 0.2 eV over the triplet minimum. At the position of the $T_1$ minimum, the computed SOC is also remarkably strong amounting to 167 cm$^{-1}$. Consistently with the similarities between the $S_1$ and $T_1$ minima, reaching the $T_1/S_0$ crossing demands the $C(NH_2)_2$ moiety to become perpendicular to the C-S bond.

The strong coupling between the $S_1$ electronic state and the triplets and the accessibility of the $S_1/S_0$ conical intersection suggest the potential competition between the deactivation exclusively along the singlet manifold and the eventual relaxation through population of triplet states. Thus, in order to provide a time resolved picture of the actual deactivation mechanism of electronically excited thiourea we have undertaken non-adiabatic molecular dynamics simulations, including both singlet and triplet states.

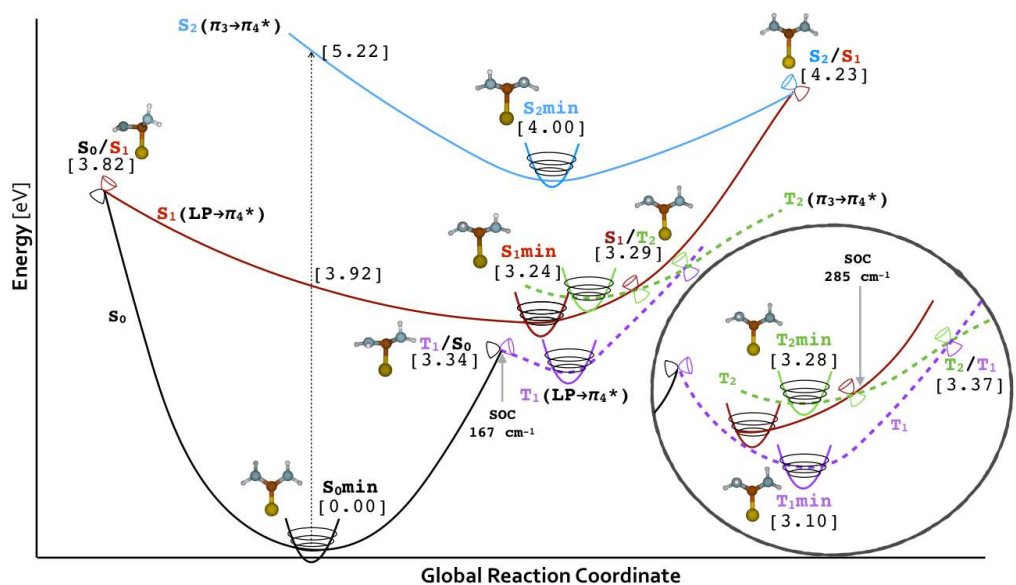

**Figure 4.-** Potential energy profile for thiourea inferred from quantum mechanical studies. All energies (in eV) were computed at the MS-CASPT2/CASSCF/(10,7)/ANO-L level of theory. Some SOC values (in cm⁻¹) are also shown.

Time-resolved thiourea's deactivation dynamics was studied both in gas phase and considering the molecule immersed in a water cluster. For this study, the first three singlets and two triplets, i.e. 9 spin-mixed electronic states in the spin-adiabatic approach, were considered.

For the gas phase relaxation dynamics, a total number of 445 trajectories were propagated for 10 ps. Figure 5a illustrates the evolution in time of the population of the spin-diabatic states. The dynamics were started from the 9-th spin-adiabatic state, which in practice corresponds to the FC $S_2$ ($\pi_3 \rightarrow \pi_4^*$). The average transition energy amounts to 3.91 eV.

The coordinate most affected by the $\pi\pi^*$ excitation is the C-S bond, in agreement with our static results (see discussion above). In the time domain comprised between 0-1 ps, when the $S_2$ carries the largest population, the C-S bond distance was found to oscillate with a period of 50 fs, see Figure 6. Simultaneous to the CS bond oscillations, we also register in phase oscillations of the two C-N bonds with a period of 22.8 fs. These oscillations progressively wash out due to the randomization of the excess of vibrational energy (IVR) and the evolution of the population to other electronic states. The most important deexcitation channel of the $S_2$ (86% of the trajectories) is by far internal conversion to the $S_1$ (a schematic population flow, in terms of forward and backward hops, is shown in figure S1). Interestingly, the maxima in the oscillation of the C-S bond length coincide with the minimum values for the $S_2$-$S_1$ energy gap within this time domain (see Figure 6), and in fact $S_2$->$S_1$ transitions take place at large C-S distances (1.83 Å in average). Again, this is in perfect agreement with our CASPT2 prediction for the $S_2/S_1$ conical intersection, which presents a very stretched C-S bond distance of 2.26 Å. These oscillations of the CS bond and, thus, in the magnitude of the $S_2$-$S_1$ energy gap are responsible for the intense recrossing detected during the 10 ps of the simulations.

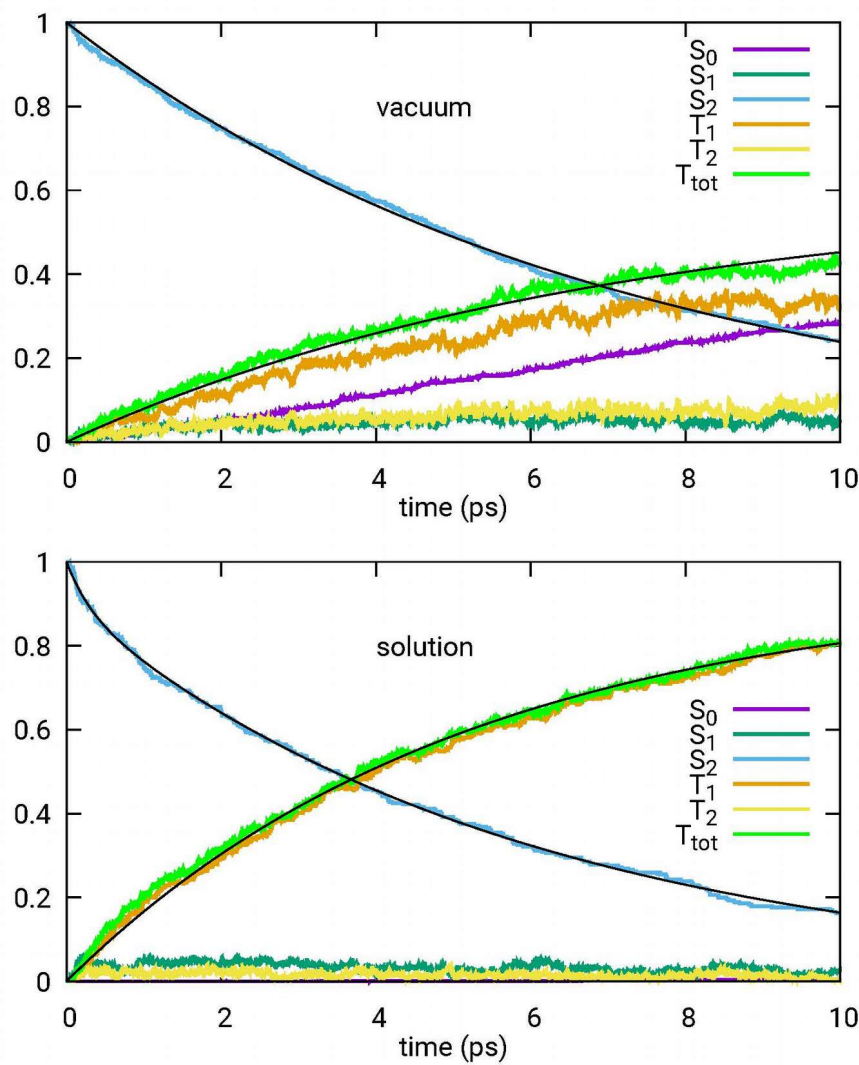

**Figure 5.-** Time evolution of the populations of the spin-diabatic states for thiourea in vacuum (upper panel) and in a cluster of water molecules (lower panel). $T_{tot}$ is the total triplet population. The black curves are mono- or bi-exponential fits of the state populations.

After an initial rising, the population of $S_1$ state remains low and constant, due to the fast transfer of population from this state to $T_2$, $T_1$ and the ground state. In fact, the $T_2$ lies always very close in energy to the $S_1$, consistently with our CASPT2 calculations which predict the $S_1$ and $T_2$ minima very close in energy. This proximity between the $S_1$ and $T_2$ translates in frequent back and forward hops between the two electronic states (ca. 2 transitions per ps per trajectory). Subsequent $T_2 \rightarrow T_1$ internal conversion would lead to the fast depopulation of the $T_2$ state in favor of the most stable triplet, $T_1$, which carries the largest population at the final time of the simulation.

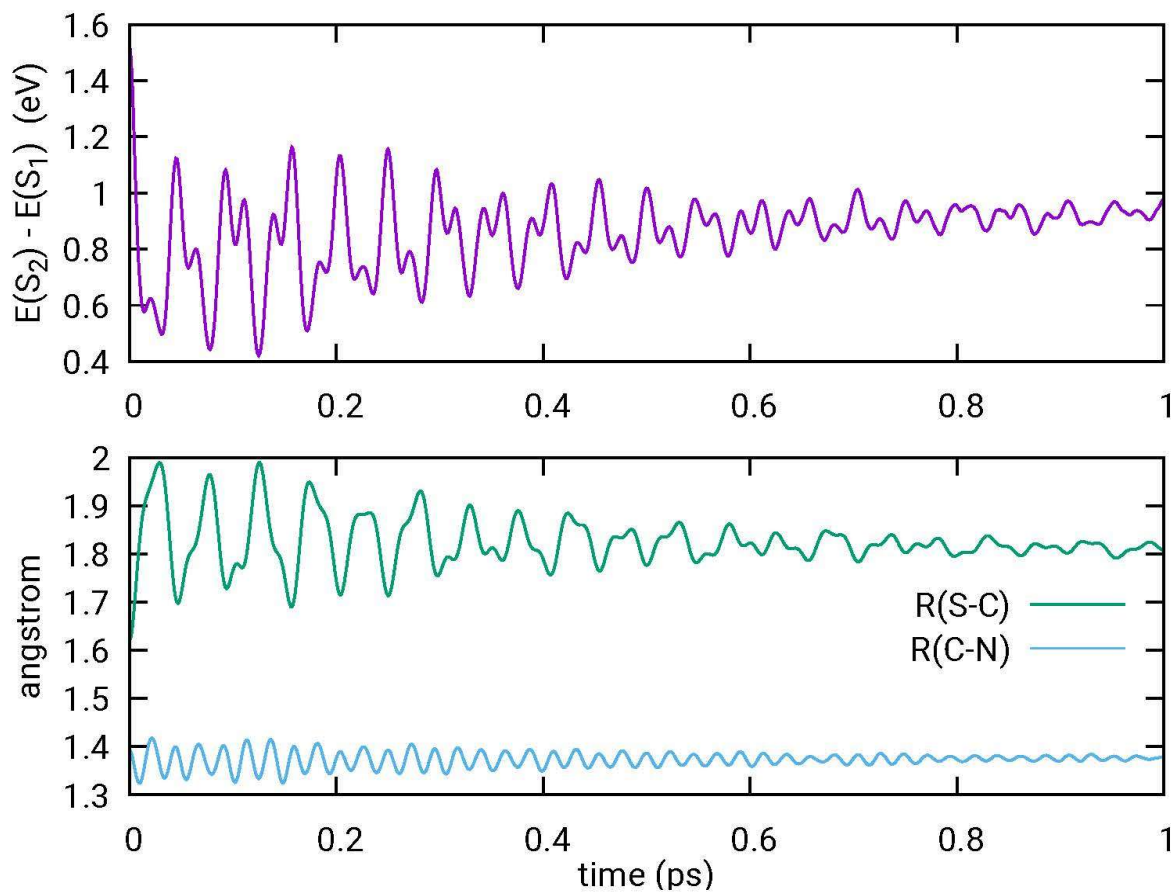

**Figure 6.-** Time evolution of selected average energetic and geometrical variables, for thiourea in vacuum. Upper panel: E(S$_2$) − E(S$_1$) energy difference, in eV. Lower panel, C-S and C-N bond distances, in Å.

According to the above considerations, the decay of the S$_2$ population (*PS$_2$*) and the rise of the total triplet population *PT* can be fitted with a single exponential with the same time constant of τ=6.98 ps and a triplet quantum yield of φ$_T$=0.59.

$$PS_2 = e^{\frac{-t}{\tau}}$$

$$PT = \phi_T \left(1 - e^{\frac{-t}{\tau}}\right)$$

At the final time of the simulations, the population of the S$_0$ is 0.28. Most of the trajectories (67%) decay to the S$_0$ from the T$_1$ or the S$_1$. Alternatively, decay to the S$_0$ occurs via the S$_2$-> S$_0$ transitions (33%). Decay to the S$_0$ takes place predominantly through the T$_1$/S$_0$ CI region, which is only 0.29 eV above the T$_1$ minimum at AM1 level of theory, in line with the CASPT2 results, which locate this crossing 0.24 eV over the energy of the $T_1$ minimum. It must be noted that the $S_1$ state lies close in energy at this region of the PES, allowing deactivation to the GS directly along the singlet manifold. A minor role is played by the $S_1/S_0$ CI, which actually corresponds to a three-state degeneracy point $S_1/T_1/S_0$ and lies higher in energy (2.70 eV vs. 2.14 eV at the AM1 level of theory) also in qualitative agreement with our CASPT2 results which predicts energies of 3.82 and 3.34 eV for these two crossings (recall Figure 4). $S_2$->$S_0$ transitions take place in the region of the $S_{2min}$, where larger energy gaps were registered (compare average gaps for $S_2$->$S_0$ transitions of 2.79 eV with 0.21 ($T_1$-> $S_0$) and 0.43 ($S_1$->$S_0$) eV). The three panels of Figure 7 show the details of a representative trajectory in vacuum. This trajectory starts from the spectroscopic state $S_2$. As for most of the trajectories examined, the system takes some picoseconds, ca 3.85 ps, to decay to the $S_1$ state, along which the oscillatory motion of the CS bond allows reducing the $S_2$-$S_1$ energy gap. For this particular trajectory, an earlier $S_2$->$S_1$ transition, ca. 1.25 ps, is registered, which is followed by fast a hop of the system backward to the initial electronic state, $S_2$. Once in the $S_1$, the proximity between the $S_1$, $T_2$ and $T_1$ potentials translates into several fast recrossings between the three electronic states. Starting from t=4.2 ps the most stable triplet $T_1$ is populated, until it undergoes an intersystem crossing to the ground state at t=4.55 ps.

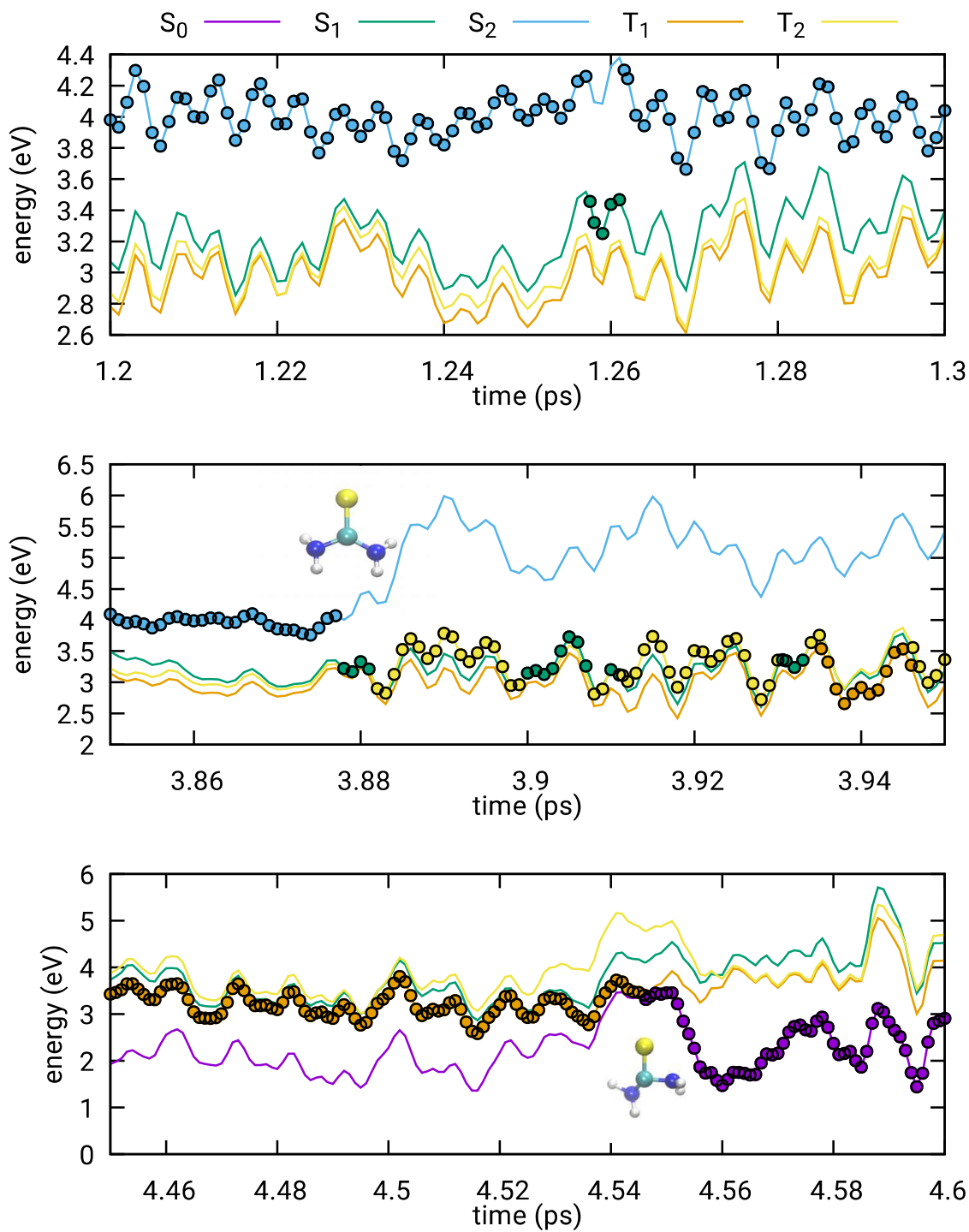

**gure 7.-** Time evolution of spin-diabatic energies for a "representative" trajectory in three selected time intervals, for thiourea in vacuum. Colored circles label the state on which the nuclear trajectory is propagated. Transitions between electronic states are located comparing the color of the circles.

In the following, we will analyze the effect of a surrounding cluster of water molecules in the decay dynamics of thiourea. As for the vacuum simulations, the water solvent dynamics were started from the $S_2$ state, with a slightly lower average excitation energy of 3.84 eV. As expected, the ultrafast (i.e. sub-ps) dynamics is not much influenced by the presence of the solvent. The in phase oscillations of the C-S bond noted above in the sub-ps time scale are also present when thiourea is immersed in solution.

Considering the transitions between the $S_2$ and the $S_1$, the main difference with respect to the gas phase dynamics is that there are many fewer backward $S_1$->$S_2$ hops (see also figure S2). In fact, these backward transitions are completely suppressed after 2.5 ps. This is ascribed to the dissipation of the excess of vibrational energy to the environment, which turns transitions to lower lying electronic states irreversible. In fact, this is reflected in the evolution of the average nuclear kinetic energy $E_k(t)$ for all the trajectories. The fit of $E_k(t)$ with a biexponential function, as described in ref [52], delivers a time constant for the thermal relaxation of thiourea in water of 6.15 ps, which is in agreement with the behavior reported above.

The most striking difference with respect to the gas phase results is the complete absence of transitions to the ground state within the first 10 ps that the simulations last, so that at the final time of the simulations the $T_1$ potential is the electronic state collecting most of the population (0.8). On the one hand, the decrease in the kinetic energy due to the thermalization reduces the Born Oppenheimer couplings. For this reason, the $S_2$->$S_0$ transitions from the $S_{2min}$ region are suppressed. On the other hand, thermalization also reduces the nuclear configurational space explored during the photodynamics. As a matter of fact, no trajectory running on $S_1$ or $T_1$ was able to access the region where the energy difference with the $S_0$ is smaller than 0.20 eV. This is in contrast with our gas phase simulations, where 86% of the trajectories running on the $T_1$ and 7% of the trajectories running on the $S_1$ were able to reach regions in which the energy gap with $S_0$ amounts to 0.2 eV or less. Interestingly, specific interactions with solvent molecules might also play a role in this respect. In particular, during the simulations hydrogen bonds are dynamically formed and broken between the $NH_2$ groups of thiourea and water molecules. These interactions might hinder the rotation and/or planarization of the $NH_2$ groups, which are important to reach the $T_1/S_0$ and $S_1/T_1/S_0$ CI regions. To have an appreciation of the hydrogen bonding interactions between thiourea and water we evaluated the NH-O and N-HO hydrogen bond energies for the $H_2O$-thiourea complex within the semiempirical QM/MM framework.

We found values in the range 4.6-6.1 kcal/mol, which are in line with the hydrogen bond strengths reported in the literature [53] for the water dimer (4.45 kcal/mol) and the water-ammonia complex (5.77 kcal/mol). We expect therefore a qualitatively correct account of the hydrogen bonding interactions in our dynamics calculations.

The suppression of the $S_1$->$S_2$ backward hops lead to an increased decay rate of the $S_2$ compared to the gas phase, at least in the first few ps. Therefore, in this case the decay of the $S_2$ is best fitted by a biexponential function, with $\omega=0.9$, $\tau_1=0.30$ ps and $\tau_2=5.87$ps

$$P\,S_2 = \omega\,e^{-t/\tau_2} + (1-\omega)\,e^{-t/\tau_1}$$

However, by fitting the decay of the $S_2$ population with a monoexponential function, we get $\tau_{solv}=5.13$ ps, which is shorter than the $S_2$ lifetime (6.98 ps) obtained in the gas phase. Using $\tau_{solv}$ to fit the rise of the triplet population *PT* we obtain a triplet quantum yield of 0.94 for thiourea in solution.

# Conclusions

By successfully combining the static mapping of the topography of the excited state potential energy surface with state of the art multireference methods and excited state molecular dynamics simulations, including spin orbit and non-adiabatic couplings, we have characterized the relaxation mechanism of photoexcited thiourea in vacuum and water environments.

Our results show that carbonyl-by-thiocarbonyl exchange shifts to lower energies the absorption spectrum, though the shift does not affect all the transitions composing the lowest lying region of the spectrum to the same extent, i.e. whilst the brightest $S_2$ electronic state shifts by 1 eV, the most stable $S_1$ stabilizes by more than 2 eV. Oxygen-by-sulfur substitution was also found to affect the topography of the excited state potential energy surfaces. In the case of urea, very fast $S_2$->$S_1$ internal conversion is expected due to the proximity of these two electronic states at the Franck-Condon region. The minimum energy path from the Franck-Condon region along the $S_2$ potential of thiourea detects, instead, a minimum that would slow down the decay to lower lying electronic states, and which translates into a lifetime for the $S_2$ amounting to ca. 7 ps. The decay mechanism from the $S_1$ is also different for urea and thiourea. Two main decay funnels were computed for urea: the $S_1$/$S_0$ funnel that would be accessible after surmounting an energy barrier of 0.3 eV and the $S_1$/$T_1$ located very close to the position of the singlet minimum. Despite for thiourea the $S_1$/$S_0$ was located 0.6 eV above the $S_1$ minimum, no barriers were located along this decay channel. Some differences were also detected for the decay from the $S_1$ along the triplet manifold in the two systems. In fact, whilst urea is expected to directly decay to the most stable triplet state from the $S_1$, in the case of thiourea, the system would preliminary decay to the $T_2$ with which a strong coupling has been calculated, that would eventually relax to the most stable triplet, $T_1$. A third minor channel, involving direct relaxation $S_2 \rightarrow S_0$ was also registered in our gas phase molecular dynamics simulations. The incorporation of solvent solute interactions to our model was found to significantly slow down the decay dynamics of the system to the ground state, through both specific interactions and thermalization effects, leading to a triplet quantum yield close to 1.

# Acknowledgments

This work has been supported by the Project CTQ2015-63997- C2 of the Ministerio de Economía y Competitividad of Spain. I.C. gratefully acknowledges the "Ramón y Cajal" program of the Ministerio de Economía y Competitividad of Spain. N. A. thanks the Marie Curie Actions, within the Innovative Training Network-European Join Doctorate in Theoretical Chemistry and Computational Modelling TCCM-ITN-EJD-642294, for financial support. Computational time from the Centro de Computación Científica (CCC) of Universidad Autónoma de Madrid is also gratefully acknowledged. G.G. acknowledges funding from the University of Pisa, PRA_2017_28.